\documentclass[submission, copyright, creativecommons, svgnames]{eptcs}

\providecommand{\event}{PROLE 2016}

\usepackage{amsmath}
\usepackage{amssymb}

\usepackage{tikz}
\usetikzlibrary{arrows}
\usepackage{xspace}
\usepackage{graphics}

\usepgflibrary{shapes.geometric}
\usepgflibrary[shapes.geometric]
\usetikzlibrary{shapes.geometric}
\usetikzlibrary[shapes.geometric]
\usetikzlibrary[matrix]
\usetikzlibrary{arrows, decorations, decorations.markings}
\usetikzlibrary{positioning}

\usepackage{listings}

\newcommand*\lstinputpath[1]{\lstset{inputpath=#1}}
\lstinputpath{code}

\usepackage{array,arydshln}

\usepackage{xcolor}

\definecolor{dkgreen}{rgb}{0,0.6,0}
\definecolor{gray}{rgb}{0.5,0.5,0.5}
\definecolor{mauve}{rgb}{0.58,0,0.82}
\definecolor{light-red}{rgb}{1.0,0.8,0.8}
\definecolor{yellow-orange}{rgb}{1.0,0.85,0.05}
\definecolor{gray-blue}{rgb}{0.57,0.72,0.84}
\definecolor{light-gray}{gray}{0.80}

\definecolor{DarkGreen}{rgb}{0.0, 0.2, 0.13}

\definecolor{apricot}{rgb}{0.98, 0.81, 0.69}
\definecolor{bananayellow}{rgb}{1.0, 0.88, 0.21}
\definecolor{babyblueeyes}{rgb}{0.63, 0.79, 0.95}

\usepackage{soul}
\sethlcolor{titlebgcolor}

\usepackage{wrapfig}
\usepackage{times} 
\usepackage{stmaryrd} 
\usepackage{centernot}

\usepackage{comment}
\usepackage{setspace}
\excludecomment{enclosedreminders}

\usepackage[
     disable,
    textwidth=0.17\textwidth,textsize=scriptsize]{todonotes}

\newcommand{\stcomm}[1]{\todo[color=red!20,bordercolor=red,linecolor=red,size=\scriptsize, caption={}]{ST: #1}}

\newcommand{\mclcomm}[1]{\todo[color=Beige,bordercolor=DarkGoldenrod,linecolor=DarkGoldenrod]{MCL: #1}}

\newcommand{\mclcommin}[1]{\todo[inline,color=Beige,bordercolor=DarkGoldenrod,linecolor=DarkGoldenrod]{MCL: #1}}
\newcommand{\jmcomm}[1]{\todo[color=blue!20,bordercolor=blue,linecolor=blue,size=\scriptsize, caption={}]{JM: #1}}

\definecolor{dkgreen}{rgb}{0,0.6,0}
\definecolor{gray}{rgb}{0.5,0.5,0.5}
\definecolor{mauve}{rgb}{0.58,0,0.82}

\lstdefinestyle{cstyle}{
  frame=tb,
  escapechar={@},
  language=c,
  aboveskip=3mm,
  belowskip=3mm,
  showstringspaces=false,
  columns=flexible,
  basicstyle={\small\ttfamily},
  numbers=none,
  numberstyle=\tiny\color{gray},
  keywordstyle=\color{blue},
  keywordstyle=[2]\color{dkgreen},
  keywordstyle=[3]\color{magenta},
  commentstyle=\color{gray},
  stringstyle=\color{mauve},
  breaklines=true,
  breakatwhitespace=true,
  tabsize=3,
}

\lstdefinestyle{hstyle}{
  frame=tb,
  escapechar={$},    
  mathescape,         
  language=haskell,
  aboveskip=3mm,
  belowskip=3mm,
  showstringspaces=false,
  columns=flexible,
  basicstyle={\small\ttfamily},
  numbers=none,
  numberstyle=\tiny\color{gray},
  keywordstyle=\color{blue},
  keywordstyle=[2]\color{magenta},
  commentstyle=\color{gray},
  stringstyle=\color{red},
  breaklines=true,
  breakatwhitespace=true,
  tabsize=3,
  morekeywords={},
  keywords=[2]{CAssign, CAddAssOp,CAddOp,CBinary,CAssignOp, CCompound,
    CFor, CLeOp, CBlockStmt, CUnary, CPostIncOp, undefNode, CExpr, CMulOp}
}

\lstdefinestyle{cstyleTable}{
  frame=tb,
  escapechar={@},
  language=c,
  aboveskip=3mm,
  belowskip=3mm,
  showstringspaces=false,
  columns=flexible,
  basicstyle={\small\ttfamily},
  numbers=none,
  numberstyle=\small\color{gray},
  keywordstyle=\color{blue},
  keywordstyle=[2]\color{dkgreen},
  keywordstyle=[3]\color{magenta},
  commentstyle=\color{gray},
  stringstyle=\color{mauve},
  breaklines=true,
  breakatwhitespace=true,
  tabsize=3,
  frame=none,
}

\lstdefinestyle{cstyleRules}{
  escapechar={@},
  language=c,
  aboveskip=3mm,
  belowskip=3mm,
  showstringspaces=false,
  columns=flexible,
  basicstyle={\small\ttfamily},
  numbers=none,
  numberstyle=\tiny\color{gray},
  keywordstyle=\color{blue},
  keywordstyle=[2]\color{dkgreen},
  keywordstyle=[3]\color{magenta},
  commentstyle=\color{gray},
  stringstyle=\color{mauve},
  breaklines=true,
  breakatwhitespace=true,
  tabsize=3,
  keywords=[2]{cexpr,bin_oper,cop,vector_space,scalar_field,contains_expr,una_oper,cstmts,cstmt,subs,no_mod,no_mod_use,is_distributive,has_expr,no_deps,new_var,ctype,replace, pure, times},
  keywords=[3]{pattern,generate,condition,metrics,assert},  
}

\lstdefinestyle{cstyleTikz}{
  escapechar={@},
  language=c,
  aboveskip=3mm,
  belowskip=3mm,
  showstringspaces=false,
  columns=flexible,
  basicstyle={\small\ttfamily},
  numbers=none,
  numberstyle=\small\color{gray},
  keywordstyle=\color{blue},
  keywordstyle=[2]\color{dkgreen},
  keywordstyle=[3]\color{magenta},
  commentstyle=\color{gray},
  stringstyle=\color{mauve},
  breaklines=true,
  breakatwhitespace=true,
  tabsize=3,
  keywords=[2]{cexpr,bin_oper,cop,vector_space,scalar_field,contains_expr,una_oper,cstmts,cstmt,subs,no_mod,no_mod_use,is_distributive,has_expr,no_deps,new_var,ctype,replace, pure, times},
  keywords=[3]{pattern,generate,condition,metrics},  
}

\newcommand{\ihaskell}[1]{\lstinline[style=hstyle,basicstyle={\small\ttfamily}]{#1}}

\newcommand{\rw}{\ensuremath{\Rightarrow}}
\newcommand{\when}{\ensuremath{\mathbf{when}}}

\newcommand{\nowrites}{\centernot{\mapsto}}
\newcommand{\noreads}{\centernot{\mapsfrom}}
\newcommand{\pure}{\mathrm{pure}}
\newcommand{\fresh}{\mathrm{fresh}}
\newcommand{\writes}{\mathrm{writes}}
\newcommand{\occurs}{\mathrm{occurs~in}}

\newcommand{\distributesover}{\mathrm{distributes\_over}}
\newcommand{\nowritesexcarrays}{\underset{_{-a[l]}}{\nowrites}}
\newcommand{\nowritesarray}{\underset{_{a[l]}}{\overset{_<}{\nowrites}}}

\usepackage{soul}
\sethlcolor{pink}

\newcommand{\stml}{\textsc{stml}\xspace}

\usepackage{stmaryrd}

	\pgfmathsetmacro\lheight{1.6}
	\pgfmathsetmacro\dwidth{1.0}
	\pgfmathsetmacro\roffset{9}
	\pgfmathsetmacro\loffset{1.5}

\title{Towards a Semantics-Aware Code Transformation Toolchain for
       Heterogeneous Systems\footnote{Work partially funded by EU
         FP7-ICT-2013.3.4 project 610686 POLCA, Comunidad de Madrid
         project S2013/ICE-2731 N-Greens Software, and MINECO Projects
         TIN2012-39391-C04-03 / TIN2012-39391-C04-04 (StrongSoft),
         TIN2013-44742-C4-1-R (CAVI-ROSE), and TIN2015-67522-C3-1-R
         (TRACES).}}  
\author{
\hspace{-3ex}\parbox{0.4\textwidth}{\centering
Salvador Tamarit \qquad Julio Mari\~no
\institute{Universidad Polit\'ecnica de Madrid\\ 
	Campus de Montegancedo 28660\\
	Boadilla del Monte, Madrid, Spain}
\email{\footnotesize\{salvador.tamarit,julio.marino\}@upm.es}}
\and
\hspace{0ex}\parbox{0.45\textwidth}{\centering
Guillermo Vigueras \qquad Manuel Carro\footnote{Manuel Carro
    is also affiliated with the Universidad Polit\'{e}cnica de Madrid}
\institute{IMDEA Software Institute \\
             Campus de Montegancedo 28223 \\
             Pozuelo de Alarc\'on, Madrid, Spain}
\email{\footnotesize\{guillermo.vigueras,manuel.carro\}@imdea.org}}
}
\def\titlerunning{Towards a Semantics-Aware Code Transformation Toolchain for
       Heterogeneous Systems}
\def\authorrunning{S. Tamarit, J. Mari\~no, G. Vigueras \& M. Carro}

\begin{document}

\pagestyle{empty} 
\maketitle

\begin{abstract}
Obtaining good performance when programming heterogeneous computing
platforms poses significant challenges. 
We present a program transformation environment, implemented in
Haskell, where architecture-agnostic scientific C code with semantic
annotations is transformed into functionally equivalent code better
suited for a given platform.  
The transformation steps are represented as rules
that can be fired when certain syntactic and 
semantic conditions are fulfilled.  These rules are not hard-wired into
the rewriting engine: they are written in a C-like language and are
automatically processed and incorporated into the rewriting engine.
That makes it possible for end-users to add their own rules or to
provide sets of rules that are adapted to certain specific domains
or purposes.
%
\end{abstract}




\section{Introduction}
\label{sec:introduction}

There is a strong trend in high-performance computing towards the
integration of heterogeneous computing elements:
vector processors, GPUs, FPGAs, etc.  Each of these components is
specially suited for some class of computations, which makes the
resulting platform able to excel in performance by mapping
computations to the unit best suited to execute them.  Such platforms
are proving to be a cost-effective alternative to more traditional
supercomputing
architectures~\cite{danalis2010-shoc-benchmark,lindtjorn2011-beyond-micro}
in terms of performance and energy consumption.  However, this
specialization comes at the price of additional hardware and, notably,
software complexity.  Developers must take care of very different
features to make the most of the underlying computing infrastructure.
Thus, programming these systems is restricted to a few experts, which
hinders its widespread adoption, increases the likelihood of bugs, and
greatly limits portability.
For these reasons, defining programming models that ease the task of
efficiently programming heterogeneous systems has become a topic of
great relevance and is the objective of many ongoing efforts (for
example, the POLCA project
{\small\url{http://polca-project.eu}}, which focuses on scientific
applications). 


Scientific applications sit at the core of many research projects of
industrial relevance that require, for example, simulating physical
systems or  numerically solving differential equations.  One
distinguishing characteristic of many scientific applications is that
they rely on a large base of existing algorithms.  These algorithms
often need to be ported to new architectures and exploit 
their computational strengths to the limit, while avoiding pitfalls and
bottlenecks.  Of course, these new versions have to preserve the
functional properties
of the original code.
%
Porting is often carried out by transforming or replacing certain fragments
of code to improve their performance in a given architecture while
preserving their semantics.
Unfortunately, (legacy) code often does not clearly spell its meaning or the
programmer's intentions, although scientific code usually
follows patterns rooted in its mathematical origin.

\begin{figure*}[t]
 \begin{center}
   \setlength\dashlinedash{1pt}
   \setlength\dashlinegap{2pt}
   \setlength\extrarowheight{-2ex} 
   \begin{tabular}{:l:l:l:}
     \hdashline
      \multicolumn{1}{:c:}{0 - original code} &
     \multicolumn{1}{:c:}{1 - \textsc{For-LoopFusion}} &
     \multicolumn{1}{:c:}{2 - \textsc{AugAdditionAssign}}
     \\\hdashline
     \begin{minipage}[t]{5cm}
     \vspace*{-1.5ex}
\begin{lstlisting}
float c[N], v[N], a, b;
@\aftergroup\blackcolor@@\hl{\textbf{for}(\textbf{int} i=0;i<N;i++)}@
  @\hl{c[i] = a*v[i];}@

@\hl{\textbf{for}(\textbf{int} i=0;i<N;i++)}@
  @\hl{c[i] += b*v[i];}@
\end{lstlisting}
     \vspace*{-1.5ex}
     \end{minipage}
     &
     \begin{minipage}[t]{4.7cm}
     \vspace*{-1.5ex}
\begin{lstlisting}
@\aftergroup\redcolor@for(int i=0;i<N;i++) {
   c[i] = a*v[i];
   @\hl{c[i] += b*v[i]}@;
}@\aftergroup\blackcolor@
\end{lstlisting}
     \vspace*{-1.5ex}
     \end{minipage}
     &
     \begin{minipage}[t]{4.9cm}
     \vspace*{-1.5ex}
\begin{lstlisting}
@\aftergroup\blackcolor@for(int i=0;i<N;i++) {
   @\hl{c[i] = a*v[i];}@
   @\aftergroup\redcolor@@\hl{c[i] = c[i] + b*v[i]}@@\aftergroup\blackcolor@@\hl{;}@
}
\end{lstlisting}
     \vspace*{-1.5ex}

     \end{minipage}
     \\\hdashline
     \multicolumn{1}{:c:}{3 - \textsc{JoinAssignments}} &
     \multicolumn{1}{:c:}{4 - \textsc{UndoDistribute}} &
     \multicolumn{1}{:c:}{5 -\textsc{LoopInvCodeMotion}}
     \\\hdashline
     \begin{minipage}[t]{5cm}
     \vspace*{-1.5ex}
\begin{lstlisting}
for(int i=0;i<N;i++)
  @\aftergroup\redcolor@c[i] = @\hl{a*v[i]+b*v[i]}@;
\end{lstlisting}
     \vspace*{-1.5ex}
     \end{minipage}
     &
     \begin{minipage}[t]{4.7cm}
     \vspace*{-1.5ex}
\begin{lstlisting}
@\aftergroup\blackcolor@for(int i=0;i<N;i++)
   @\hl{c[i] = }@@\aftergroup\redcolor@@\hl{(a+b) * v[i]}@@\aftergroup\blackcolor@@\hl{;}@
\end{lstlisting}
     \vspace*{-1.5ex}
     \end{minipage}
     &
     \begin{minipage}[t]{4.9cm}
     \vspace*{-1.5ex}
\begin{lstlisting}
@\aftergroup\redcolor@float k = a + b;@\aftergroup\blackcolor@
for(int i=0;i<N;i++)
   @\aftergroup\redcolor@c[i] = k * v[i];@\aftergroup\blackcolor@
\end{lstlisting}
     \vspace*{-1.5ex}
     \end{minipage}
     \\\hdashline
   \end{tabular}
 \end{center}
 \caption{A sequence of transformations of a piece of C code to compute
          $\textbf{c}=a\textbf{v}+b\textbf{v}$.} 
 \label{fig:code_trans_seq}
\end{figure*}

Our goal is to obtain a framework for the transformation of
(scientific), architecture-agnostic C code.  The framework should be
able to transform existing code into a functionally equivalent program,
only better suited for a given platform.  Despite the broad range of
compilation and refactoring tools
available~\cite{Bagge03CodeBoost,visser04:stratego-XT-0.9,Schupp2002},
no existing tool fits our needs by being adaptable enough to flexibly recognize
specific source patterns and generate code better adapted to different
architectures
(Section~\ref{sec:related}),  
so we decided to implement our own transformation framework. 
Its core is a code rewriting engine,
written in Haskell, that works at the \emph{abstract syntax tree} (AST) 
level.  The engine executes transformation rules written in a
C-like, domain-specific language (\stml, inspired by
CTT~\cite{Boekhold1999} and CML~\cite{Brown2005-tr-opt_trans_hw}).
This makes understanding the meaning of the rules and defining
additional rulesets for specific domains or targets easy for C
programmers.  
The tool does not have hard-wired strategies to select which rules are
the most appropriate for each case.  Instead, it is modularly designed
to use external oracles that help in selecting which rules have to be
applied.  In this respect, we are developing human interfaces and
machine learning-based tools that advice on
the selection of the most promising transformation
chain(s)~\cite{vigueras16:learning-prole}.  
The tool also includes an interactive mode to allow for more steering
by expert users.  When code deemed adequate for the target
architecture is reached, it is 
handed out to a \emph{translator} in charge of adapting it to the
programming model of the target platform.

Fig.~\ref{fig:code_trans_seq} shows a sample code
transformation sequence,
containing an original fragment of C code along with the result of
stepwise applying a number of transformations. Although the examples
presented in this paper are simple, the tool is able to transform
much more complex, real-life code, including code with
arbitrarily nested loops (both for collapsing them in a single loop and creating
them from non-nested loops), code that needs inlining, and others.
Some of these transformations are currently done by existing
optimizing compilers.  However, they are usually performed internally,
at the \emph{intermediate representation} (IR) level, and with few,
if any, opportunities for user intervention or tailoring, which falls
short to cater for many relevant situations that we want to address:

\begin{itemize}
\item Most compilers are designed to work with minimal (if any)
  interaction with the environment.  While this situation is optimal when it can
  be applied, in many cases static analysis cannot discover the
  underlying properties that a programmer knows.  For example, in
  Fig.~\ref{fig:code_trans_seq}, a compiler would rely on native
  knowledge of the properties of multiplication and addition.
  If these operations were substituted by calls to functions
  implementing operations with the same properties (distributivity,
  associativity, commutativity), such as operations on matrices, the
  transformation presented would be feasible but unlikely to be
  performed by a compiler relying solely on static analysis.

\item Most compilers have a set of \emph{standard} transformations
  that are useful for usual architectures ---commonly
  Von Neumann-based CPU architectures.  However, when CPU-generic
  code is to be adapted for a specific architecture (e.g., FPGA,
  GPGPU) the transformations to be made 
  are not trivial and fall outside those usually implemented in
  standard compilers.  Even more, compilers (such as
  ROCCC~\cite{roccc-manual}) that accept a subset of the C language
  and generate executables or lower-level code for a specific
  architecture, need the input code to follow specific coding
  patterns, which our tool can help generate.

\item Transformations to generate code amenable to be compiled down to
  some hybrid architecture can be sometimes complex and are better
  expressed at a higher level rather than inside a compiler's
  architecture.  That could require users to come up with
  transformations that are better suited for a given coding style or
  application domain.  Therefore, giving programmers the possibility
  of defining transformations at a higher level as plugins for a
  compiler greatly enlarges the set of scenarios where automatic
  program manipulation can be applied.
\end{itemize}


\begin{figure*}
 \begin{center}
\scalebox{0.85}{
\begin{tikzpicture}[
    place1/.style={circle,draw=blue!50,fill=blue!20,thick},
    place2/.style={circle,draw=blue!50,fill=green!80,thick},
    place3/.style={circle,draw=blue!50,fill=red!40,thick},
    thick,->]
    \node (gpu) at (4.5,1.5) [place3,label=right:GPGPU (\textit{OpenCL}),label={[yshift=0.1cm,xshift=0.5cm]above:\parbox{7em}{\textcolor{red!80}{Translated code}}}] {};
    \node (omp) at (4.5,0.5) [place3,label=right:OpenMP] {};
    \node (mpi) at (4.5,-0.5) [place3,label=right:MPI] {};
    \node (fpga) at (4.5,-1.5) [place3,label=right:FPGA (\emph{\textit{MaxJ}, \textit{POROTO}})] {};
    \node (dsp) at (4.5,-2.5) [place3,label=right:DSP (\emph{\textit{FlexaWare}})] {};    
    \node (rgpu) at (2,1.5)
    [place2,label=above:\parbox{5em}{\textcolor{DarkGreen!80}{\emph{Ready}
        code}}] {} edge (gpu);
    \node (romp) at (2,0.5) [place2] {} edge (omp);
    \node (rmpi) at (2,-0.5) [place2] {} edge (mpi);
    \node (rfpga) at (2,-1.5) [place2] {} edge (fpga);
    \node (rdsp) at (2,-2.5) [place2] {} edge (dsp);    
    \node (original) at (-2.0,0)
    [place1,label=above:\parbox{4em}{\textcolor{blue!60}{Initial
        code}}] {} edge (rgpu) edge (romp) edge (rmpi) edge (rfpga) edge (rdsp);
    \draw[-] (-1.6,-3) -- (-1.6,-3.25) -- (1.95, -3.25) -- (1.95, -3);
    \draw[-] (2.05,-3) -- (2.05,-3.25) -- (4.6, -3.25) -- (4.6, -3);
    \node [blue] at (0,-3.75) {Transformation};
    \node [blue] at (3.3,-3.75) {Translation};
\node (gears) at (5.8,-5) {\includegraphics[width=3cm]{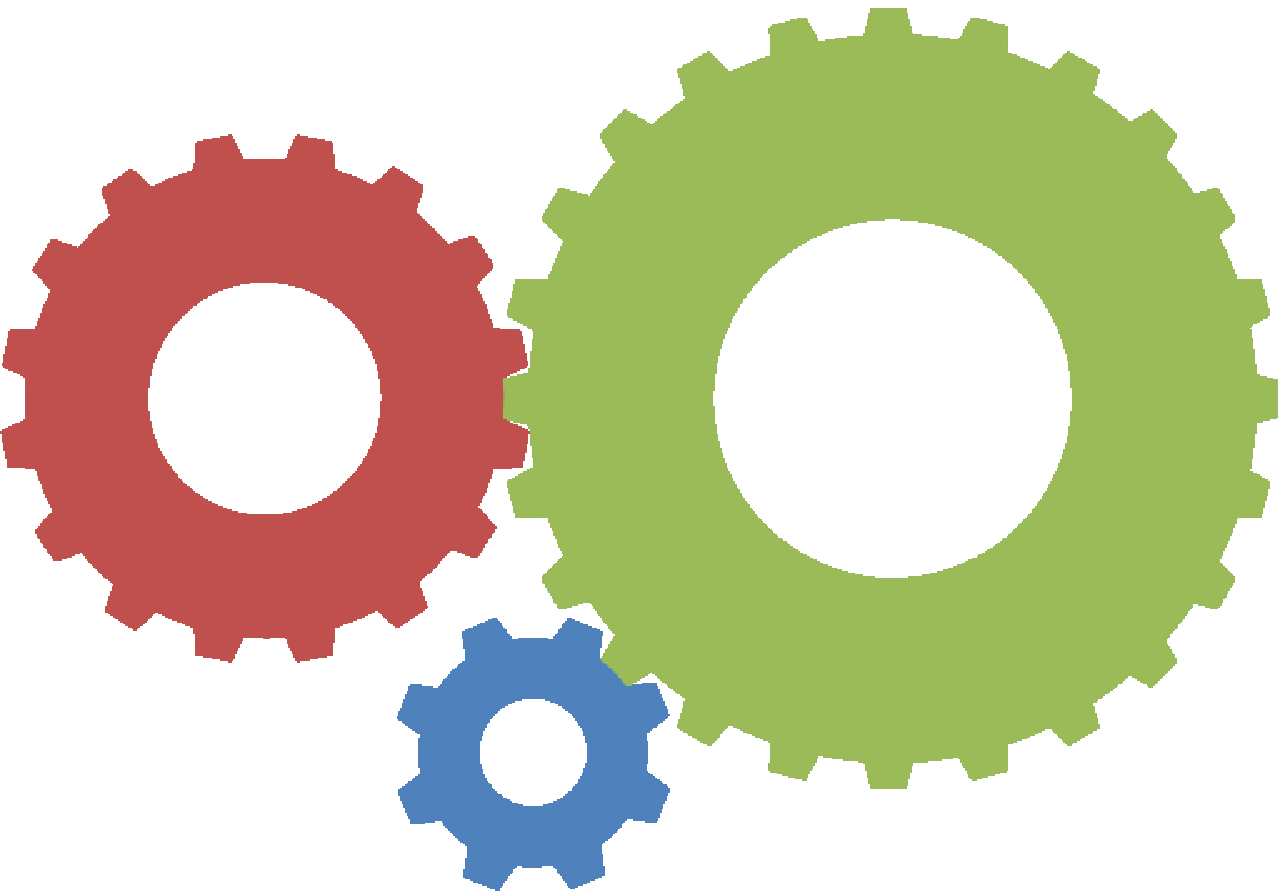}};
\node[blue,left=of gears,xshift=1.2cm] (engine) {Engine written in Haskell};
\draw[->] (-1.8, -4.4) -- (5.6, -4.4);

\node[blue,left=of original,xshift=-0.2cm,yshift=0.5cm] (rulelib)
    {\parbox{4em}{Rule  library (\stml)}};
\node[below of=rulelib,yshift=-1em] (r1)
    {\includegraphics[width=0.7cm]{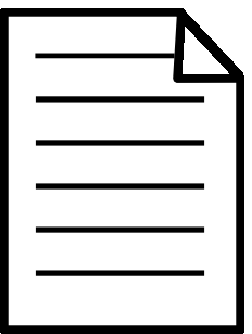}};
\node[below of=r1] (r2)
    {\includegraphics[width=0.7cm]{figures/paper.eps}};
\node[below of=r2] (r3)
    {\includegraphics[width=0.7cm]{figures/paper.eps}};

\node[blue,below of=engine,left=of engine,xshift=-0.5cm] (rulelibh)
    {\parbox{5em}{Rule library (Haskell)}};
\node[right=of rulelibh,xshift=-1cm] (h1)
    {\includegraphics[width=0.7cm]{figures/paper}
     \hspace*{-1.7em}\raisebox{1.2em}{\large\textbf{.hs}}};
\node[right=of h1,xshift=-1cm] (h2)
    {\includegraphics[width=0.7cm]{figures/paper}
     \hspace*{-1.7em}\raisebox{1.2em}{\large\textbf{.hs}}};
\node[right=of h2,xshift=-1cm] (h3) 
    {\includegraphics[width=0.7cm]{figures/paper}
     \hspace*{-1.7em}\raisebox{1.2em}{\large\textbf{.hs}}};

\draw[->] (r3) -- (h1);
\draw[->] (r2) -- (h2);
\draw[->] (r1) -- (h3);
\draw[->] (h1) -- (engine);
\draw[->] (h2) -- (engine);
\draw[->] (h3) -- (engine);

\node[blue] (rule) [right=of original,yshift=2.8cm,xshift=-1cm]
     {Rule execution(s)};
\path[-o] (rule.south) edge node[left] {\includegraphics[width=0.7cm]{figures/paper.eps}} (0,0.1);
\path[->] (engine.west) edge [bend left=70]  (rule.west);
  \end{tikzpicture}}
  \end{center}
  \caption{Architecture of the transformation tool.}
  \label{fig:ana-trans-tool}
\end{figure*}

Fig.~\ref{fig:ana-trans-tool} shows an overview of the tool, designed
to work in two stages: a \emph{transformation} phase
(Section~\ref{sec:s2stransformation}) and a \emph{translation} phase
(Section~\ref{sec:translation}).  The transformation phase rewrites
the original input code into a form that follows coding patterns
closer to what is appropriate for the destination architecture.
That code can be given to compilers that accept C code adapted to
the targeted architecture~\cite{roccc-manual}.  Additionally, this
transformation phase can be used to other purposes, such as
sophisticated code refactoring.  
The translation phase converts transformed code into code that can be
compiled for the appropriate architecture by tools that do not accept
(sequential) C code. For example, in our case
MaxJ~\cite{techologies16:max_compiler} code can be generated, as well
as C code with OpenMP annotations or with MPI calls.



%
Our efforts have focused so far on the transformation
phase~\cite{tamarit15:padl-haskell_transformation}. Our initial work
on the translation phase shows encouraging results and points to next
steps which we present in more detail in
Section~\ref{sec:conclusions}.



\section{Related Work}
\label{sec:related}



Some related approaches generate code from
a mathematical model (automatic code synthesis),
while others use (mathematical) properties to transform existing
code.  The former  
can in many cases generate underperforming code because of
its generality.  The latter usually requires that the initial code is
in some ``canonical'' form.

An example of code generation based on mathematical specifications
is~\cite{Franchetti2006}, which focuses on synthesizing code for
matrix operations.  The starting point is a mathematical formula that
is transformed (not automatically) using rewriting rules to generate
another formula that can be implemented in a hopefully more efficient
way.  This kind of approaches are often very domain-dependent and
restricted to a certain kind of formulas or mathematical objects,
which makes their application to general domains not straightforward,
if possible at all. Given that the starting point is a mathematical
formula, applying them to legacy code is not easy.  Also, their code
generation is usually based on composing blocks that correspond to
patterns for the basic formulas, where inter-optimization is often not
exercised.

\begin{enclosedreminders}
  \mclcomm{It's not clear that this makes a lot of sense here, but
    OTOH I think it's something that applies to all of the related
    work, given that many of our ideas are already elsewhere --- but
    we don't have any example with metrics, i.e., we don't show
    anything we can say is original.}
\end{enclosedreminders}
%


\begin{enclosedreminders}
  \mclcommin{Other points are how do Stratego and others deal with
    metrics (if they do), properties that are decided using external
    tools, types of transformations (e.g. does it do procedural
    transformations involving destructive assignment --- but I guess
    the answer is yes).}  \mclcommin{One more thing: (semantic)
    annotations / hints, of course}
\end{enclosedreminders}

There are some language-independent transformation tools that share
some similarities with our approach. The most relevant ones are
Stratego-XT~\cite{visser04:stratego-XT-0.9}, 
TXL~\cite{Cordy2006txl}, DMS~\cite{baxter2004dms} and
Rascal~\cite{klint2009rascal}. Stratego-XT is more oriented to
strategies than to rewriting rules, and its rule language is too
limited for our needs (e.g., rule firing does not depend on semantic conditions
that express when applying a rule is sound).  This may be adequate for a
functional language featuring referential transparency, but not
for a procedural language.
Besides, it is not designed to add analyzers as plugins,
it does not support pragmas or keeps a
symbol table.  This means that, for some cases, it is not possible to
decide whether a given rule can be soundly applied.  The last two
disadvantages are shared with TXL.  DMS is a 
powerful, industrial transformation tool that is not free and there is
not too much information of how it works internally; its overall
open documentation is scarce.  Since it transforms programs by applying
rules until a fix point is reached, the rules should be carefully
defined to ensure that they do not produce loops in the rewriting stage.
Finally, Rascal is still in alpha state and only available as binary,
and the source code is not freely and immediately accessible.

CodeBoost~\cite{Bagge03CodeBoost}, built on top of
Stratego-XT, performs domain-specific
optimizations to C++ code
following an approach similar in spirit to our proposal.  User-defined
rules specify domain-specific optimizations; code annotations are used
as preconditions and inserted as postconditions during the rewriting
process. However, it is a mostly abandoned project that, additionally,
mixes C++, the Stratego-XT language, and their rule language.  All
together, this makes it to have a steep learning curve.
\begin{enclosedreminders}
\mclcomm{Is metrics the only difference?  How about strategies?}
\end{enclosedreminders}
Concept-based frameworks such as Simplicissimus~\cite{Schupp2002}
transform C++ based on user-provided algebraic properties.  The rule
application strategy can be guided by the cost of the resulting
operation.  This cost is defined at the expression level (and not at
the statement level), which makes its applicability limited.  Besides,
their cost is defined using ``arbiters'' that do not have a global
view of the transformation process, which makes it possible to become
trapped in local minima (Sec.~\ref{sec:oracle-selection}).
Handel-C~\cite{Brown2005-tr-opt_trans_hw} performs program
transformations to optimize parallelism in its compilation into a
FPGA.  It is however focused on a subset of C enriched with extensions
to express synchronicity, and therefore some of its assumptions are
not valid in more general settings.



Other systems lay between both approaches.  They use a high-level
(declarative) language with some syntactical / semantic restrictions
that is compiled down to a lower level
formalism~\cite{DBLP:conf/dsd/BaaijKKBG10,dubach2012compiling}.  While
successful for their goals, they cannot directly tackle the problem of
adapting existing code.

Most compilers have an internal phase to perform code transformation,
commonly at the IR level.  Among the well-known open-source compilers,
CLang / LLVM probably has the better designed libraries / APIs to
perform program manipulation.
However, they were designed for compilation instead of for
source-to-source program transformation.
We tried using them but found that they are neither easy to use nor
effective in many situations.
Moreover, the design documents warn that the interface can not be
assumed to be stable.  Additionally, code transformation routines had
to be coded in C++, which made these routines verbose and full of low-level
details, and writing them error-prone.  Compiling rules to C++ is of
course an option, but the conceptual distance between the rules and
the (unstable) code manipulation API was quite large.  That pointed to
a difficult compilation stage that would need considerable maintenance.
%
Even in that case, the whole CLang project would have to be recompiled after
introducing new rules, which would make project development and testing
cumbersome, and would make adding user-defined rules complicated.

\section{Source-to-Source Transformations}
\label{sec:s2stransformation}

The code transformation tool has two main components: a parser that
reads the input program and
the transformation rules and builds an AST using Haskell data types
and  translates the rules into Haskell for faster execution,
and an engine that performs source-to-source C code transformation using
these rules.

\begin{table}[h!]
\centering
\begin{tabular}[t]{|l|}
\hline  
\begin{lstlisting}[basicstyle = \ttfamily\small]
#pragma polca scanl F INI v w
\end{lstlisting}\\\hline
\begin{lstlisting}[basicstyle = \ttfamily\small]
#pragma stml reads output(INI)
#pragma stml reads v in {0}
#pragma stml reads w in {0}
#pragma stml writes w in {1}
#pragma stml pure F
#pragma stml iteration_space 0 length(v)
\end{lstlisting}\\\hline 

\multicolumn{1}{c}{~}\\\hline  
\begin{lstlisting}[basicstyle = \ttfamily\small]
#pragma polca zipWith F v w z
\end{lstlisting}\\\hline
\begin{lstlisting}[basicstyle = \ttfamily\small]
#pragma stml reads v in {0}
#pragma stml reads w in {0}
#pragma stml writes z in {0}
#pragma stml same_length v w
#pragma stml same_length v z
#pragma stml pure F
#pragma stml iteration_space 0 length(v)
#pragma stml iteration_independent
\end{lstlisting}\\\hline

\multicolumn{1}{c}{~}\\\hline  


\begin{lstlisting}[basicstyle = \ttfamily\small]
#pragma polca map F v w
\end{lstlisting}\\
\hline
\begin{lstlisting}[basicstyle = \ttfamily\small]
#pragma stml reads v in {0}
#pragma stml writes w in {0}
#pragma stml same_length v w
#pragma stml pure F
#pragma stml iteration_space 0 length(v)
#pragma stml iteration_independent
\end{lstlisting}\\\hline

\multicolumn{1}{c}{~}\\\hline 
\begin{lstlisting}[basicstyle = \ttfamily\small]
#pragma polca fold F INI v a
\end{lstlisting}\\\hline
\begin{lstlisting}[basicstyle = \ttfamily\small]
#pragma stml reads v in {0} 
#pragma stml reads output(INI)
#pragma stml writes a
#pragma stml pure F
#pragma stml iteration_space 0 length(v)
\end{lstlisting}\\
\hline


\end{tabular}
\caption{Annotations used in the POLCA project and their translation into \stml annotations.}
\label{tab:polca_stml_translation}
\end{table}

The transformation rules
contain patterns that have to syntactically match input code and
describe the skeleton of the code to generate.  They can specify, if
necessary, conditions to ensure that their application is sound.
These conditions are checked using a combination of static
analysis and user-provided annotations (\emph{pragmas}) in the source
code with which the programmer provides additional information.
The annotations can capture properties at two different levels:
high-level properties that describe algorithmic structures and
low-level properties that describe details of the procedural code.
%
%
The decision of whether to apply a given rule depends on
two main factors:

\begin{itemize}
\item Its application must be sound.  This can be checked with the AST
in simple cases.
  Otherwise, whether a rule is applicable or not can be decided based on
  information inferred from annotations in the source code.  These annotations
  may come from external tools, such as static analyzers,
    or
  be provided by a programmer.
\item 
  The transformation should (eventually) improve some efficiency
  metric, which is far from trivial.  An interactive mode that leaves
  this decision to a final user is available.  While this is useful in
  many cases, including debugging, it is clearly not scalable and
  deciding the best rule is often difficult.  As a solution, we are working
  on a machine learning-based
  oracle~\cite{vigueras16:learning-prole} that decides
  which rule to apply based on estimations of the expected performance
  of rule application chains.



\end{itemize}

We present now the code annotations
and the rule language.  We will close this section with a description
of the interaction between the transformation tool and external
oracles.



\subsection{High-Level Annotations}
\label{sec:high_level_ann}

\begin{figure} 
  \centering
  \begin{tabular}{cc}
\begin{minipage}[b]{0.45\linewidth}
\begin{lstlisting}[basicstyle = \ttfamily\normalsize,,
                   caption=,
                   label=]
float c[N], v[N], a, b;

#pragma polca map BODY1 v c
for(int i=0;i<N;i++)
#pragma polca def BODY1
#pragma polca input v[i]
#pragma polca output c[i]
   c[i] = a*v[i];
\end{lstlisting}
\end{minipage}
&
\begin{minipage}[b]{0.55\linewidth}
\begin{lstlisting}[basicstyle = \ttfamily\normalsize]
#pragma polca zipWith BODY2 v c c
for(int i=0;i<N;i++)
#pragma polca def BODY2
#pragma polca input v[i]
#pragma polca input c[i]
#pragma polca output c[i]
   c[i] += b*v[i];
\end{lstlisting} 
\end{minipage}
  \end{tabular}
  
 \caption{Annotations for the code in Fig.~\ref{fig:code_trans_seq}.}
\label{lst:polca_ann_example}
\end{figure}

Annotations
describing semantic features of the code make it possible to capture
algorithmic skeletons at a higher level of abstraction and, at the
same time,  to express properties of the underlying code.  Our annotations
follow a functional programming style.  
For instance, \texttt{for} loops expressing a mapping between an input
and an output array are annotated with a \texttt{map} pragma such as
\texttt{\#pragma polca map F v w}.  This annotation would indicate that the loop
traverses the input array \texttt{v} and applies function \texttt{F}
to each element in \texttt{v} giving as result the array \texttt{w}.
For the annotation to be correct, we \emph{assume} that \texttt{F}
is pure, that \texttt{v} and \texttt{w} have the same length, and that
every element in \texttt{w} is computed only from the corresponding
element in \texttt{v}.  As a design decision, we do not check for
these properties, but we expect them to hold.%
\footnote{However, if some available analysis infers information
  contradicting any of these assumptions, we warn the user.}


The top boxes of the frames in
Table~\ref{tab:polca_stml_translation} list some high-level
annotations. 
For illustrative purposes, Fig.~\ref{lst:polca_ann_example} shows
an annotated version of the code in Fig.~\ref{fig:code_trans_seq}.
The listing shows how the algorithmic skeletons are parametric and
their functional parameters are obtained from blocks of C code by
specifying their formal inputs and outputs. 

\subsection{STML Properties}
\label{sec:stml_ann}

The transformation tool 
requires 
that some low-level, language-dependent properties hold to ensure
that transformations are sound.
While some of these properties can be inferred from a high-level annotation,
some of them can go beyond what can be expressed in the high-level
functional specifications. 
%
%
For example, a purely functional
semantics 
featuring referential transparency cannot capture some aspects of
imperative languages such as destructive assignment or aliasing.
\mclcomm{But we already have the possibility of an output variable
  being the same as an input variable?}
\jmcomm{Yes, that is a good point. That sentence come from the ``age
  of purity''. However, it can be argued that the inout parameters
  offer just a limited form of destructive assignment.} 
%
In our framework, these properties can be expressed in a language we
have termed \stml (from \emph{Semantic Transformation Meta-Language})
that can be used both in the code annotations and in the conditions
of the transformation rules.



\subsubsection{Syntax and Semantics of \stml Annotations}
\label{sec:stml-annotations}


\begin{lstlisting}[basicstyle=\small\ttfamily,
    float,caption=BNF grammar for \stml.,label=lst:stml_grammar]
<code_prop_list> ::= "#pragma stml" <code_prop> | 
                            "#pragma stml" <code_prop> <code_prop_list>
<code_prop>      ::= <loop_prop> | <exp_prop> <exp> | [<op>] <op_prop> <op>  | 
                     "write("<exp>") =" <location_list> | 
                     "same_length" <exp> <exp>  | "output("<exp>")"  | 
                     <mem_access> <exp> ["in" <offset_list>]
<loop_prop>      ::= "iteration_independent" | 
                     "iteration_space" <parameter> <parameter>
<exp_prop>       ::= "appears" | "pure" | "is_identity"
<op_prop>        ::= "commutative" | "associative" | "distributes_over"  
<mem_access>     ::= "writes" | "reads" | "rw"
<location_list>  ::= "{" <c_location> {"," <c_location>} "}" 
<offset_list>    ::= "{" <INT> {"," <INT>} "}"
<exp>            ::= <C_EXP>  | <C_VAR> | <polca_var_id>
<op>             ::= <C_OP> | <C_VAR> | <polca_var_id>
<c_location>     ::= <C_VAR> | <C_VAR>("["<C_EXP>"]")+
<parameter>      ::= <c_location> | <polca_var_id> | <INT>
\end{lstlisting}

Listing~\ref{lst:stml_grammar} shows 
the grammar for \stml annotations.
An intuitive explanation of its
semantics follows.

\begin{itemize}
\item \lstinline{<code_prop>} refers to code properties
  expressed through \stml annotations. 

\item{\lstinline{[<exp>] <exp_prop> <exp>}:} 
  \lstinline{<exp_prop>} denotes properties about code expressions of
  the statement immediately below the annotation. Some examples are:

\begin{itemize}
\item{\lstinline{appears <exp>}}: there is at least one occurrence of
  \lstinline{<exp>} in the statement below.

\item{\lstinline{pure <exp>}}: expression \lstinline{<exp>} is pure,
  i.e.\ it has neither side effects nor writes on any memory location.

\item{\lstinline{is_identity <exp>}}: \lstinline{<exp>} is
  an identity element. 
  High-level annotations that define the group or field in which
  \lstinline{<exp>} is the identity element must have appeared before.
\end{itemize}

\item{\lstinline{[<op>] <op_prop> <op>}:} \lstinline{<op_prop>} is an
  operator property (maybe binary). Some examples are:

\begin{itemize}
\item{\lstinline{commutative <op>}}: \lstinline{<op>} has
  the commutative property:  if \lstinline{<op>} = $f$, then $\forall x, y .~ f(x,y) = f(y,x)$.


\item{\lstinline{associative <op>}}: \lstinline{<op>} has
  the associative property: if \lstinline{<op>} = $f$, then
  $\forall x, y, z .$\\$~ f(f(x,y),z) = f(x,f(y,z))$.

\item{\lstinline{<op> distributes_over <op>}}: The first operator
  distributes over the second operator: if the operators are $f$ and
  $g$, then 
  $\forall x, y, z .~ g(f(x,y),z) = f(g(x,z),g(y,z))$.
\end{itemize}

\item{\lstinline{"write("<exp>") =" <location_list>}}: 
  the list of memory locations written on by expression
  \lstinline{<exp>} is \lstinline{<location_list>}, a list of
  variables (scalar or array type) in the C code. For example,
  \lstinline|write(c = a + 3) = {c}| and%
 ~\lstinline|write(c[i++] = a +  3) = {c[i], i}| 

\item{\lstinline{<mem_access> <exp> ["in" <offset_list>]}}: 
  \lstinline{<mem_access>} states properties about the memory accesses
  made by the statement(s) that immediately follow the
  expression \lstinline{<exp>}.  When \lstinline{<exp>} is an array,
  \lstinline{"in" <offset_list>} can state the list of positions
  accessed for reading from or writing to (depending on
  \lstinline{<mem_access>}) the array. 
  Some examples are:
 
 \begin{itemize}
 \item{\lstinline{writes <exp>}}: the set of statements associated to
   the annotation writing into a location identified by
   \lstinline{<exp>}.

 \item{\lstinline{writes <exp> "in" <offset_list>}}: this annotation
   is similar to the previous one, but for non-scalar variables within
   loops.  It specifies that for each \texttt{i}-th iteration of the
   loop, an array identified by \lstinline{<exp>} is written to in the
   locations whose offset with respect to the index of the loop is
   contained in
   \lstinline{<offset_list>}. 
   For example,

  \medskip   

\noindent
\begin{minipage}[t]{0.44\linewidth}
\begin{lstlisting}[style=cstyle]
#pragma stml writes c in {0}
for (i = 0; i < N; i++)
    c[i] = i*2;
@\\@
\end{lstlisting}
\end{minipage}
\qquad
\begin{minipage}[t]{0.5\linewidth}
\begin{lstlisting}[style=cstyle]
#pragma stml writes c in {-1,0}
for (i = 1; i < N; i++){
    c[i-1] = i;
    c[i]    = c[i-1] * 2;}        
\end{lstlisting}
\end{minipage}

\item{\lstinline{reads <exp>}}:  the set of statements associated with
  the annotation read from location \lstinline{<exp>}. 

\item{\lstinline{reads <exp> "in" <offset_list>t}}: similar to
  \lstinline{writes <exp> "in" <offset_list>} but for reading instead
  of writing.  An example follows: 

  \medskip


\noindent
  \begin{minipage}{0.44\linewidth}
\begin{lstlisting}[style=cstyle]
#pragma stml reads c in {0}
for (i = 0; i < N; i++)
    a += c[i];
\end{lstlisting}
  \end{minipage}
\qquad
  \begin{minipage}{0.5\linewidth}
\begin{lstlisting}[style=cstyle]
#pragma stml reads c in {-1,0,+1}
for (i = 1; i < N - 1; i++)
    a += c[i-1]+c[i+1]-2*c[i];    
\end{lstlisting}
  \end{minipage}

\item{\lstinline{rw <exp>}}: the set of statements associated to the
  \stml annotation reads and writes from / to location
  \lstinline{<exp>}.
\item{\lstinline{rw <exp> "in" <offset_list>}}: similar to
  \lstinline{writes <exp> "in" <offset_list>} but for reading or
  writing.
\end{itemize}
 
\item{\lstinline{<loop_prop>}}: this term represents annotations
  related with loop properties:
 \begin{itemize}
 \item{\lstinline{"iteration_space" <parameter> <parameter>}}: this
   annotation states the iteration space limits of the \texttt{for} loop
   associated with the annotation. An example would be:

     \medskip
\noindent
\begin{minipage}{0.55\linewidth}
\begin{lstlisting}[style=cstyle]
#pragma stml iteration_space 0 N-1
for (i = 0; i < N; i++)
    c[i] = i*2;
\end{lstlisting}
\end{minipage}

\item{\lstinline{"iteration_independent"}}: this annotation is used to
  state that there is no loop-carried dependencies in the body of the
  loop associated to this annotation. \mclcomm{Should it not be ``ivdep''?}
\end{itemize}

\item{\lstinline{"same_length" <exp> <exp>}:} the two C arrays given
  as parameters have the same length.

\item{\lstinline{"output("<exp>")"}:} \lstinline{<exp>} is the output
  of a block of code.

\end{itemize}

\subsubsection{Translation from High-Level to \stml 
Annotations} 
\label{sec:polca-stml-translation}

As mentioned before, annotated code is assumed to follow the semantics
given by the annotations. \mclcomm{We never clarified the semantics of
  the polca pragmas in Table~\ref{tab:polca_stml_translation}}\stcomm{Yes, we didn't. However, what I read here is that given a piece of code annotated with some pragma, this piece of code follow the semantics indicated by the pragma (whatever it is, i.e. in general)}
%
%
Using this interpretation, lower-level \stml properties can be
inferred for annotated code and used to decide which
transformations are applicable.
For example, let us consider the loop annotated with \texttt{map BODY1 v c} in
Fig.~\ref{lst:polca_ann_example}. 
In this context the assumption is that:

\begin{itemize}
\item \texttt{BODY1} behaves as if it had no side effects.  It may read
  and write from/to a global variable, but it should behave as if this
  variable did not implement a state for \texttt{BODY1}. For example, it
  may always write to a global variable and then read from it, and the
  behavior of other code should not depend on the contents of this
  variable. \mclcomm{I am not sure this makes sense.  If it has no
    side effects, different calls to \texttt{BODY1} could execute in
    parallel.  But if it writes and reads form a global variable,
    parallel execution is not safe.}\stcomm{It depends on if there is
    an implicit protection for these cases in the target languages,
    does it?}
  \jmcomm{I would not annotate it as a map in the first place if a
    global variable is modified, but it is true that nothing in the
    POLCA language disallows it. I would rather not shed focus on
    these inconsistencies... At least not here.}
\item \texttt{v} and \texttt{c} are arrays of the same size.
\item For every element of \texttt{c}, the element in the $i$-th
  position is computed by applying \texttt{BODY1} to the element in the
  $i$-th position of \texttt{v}.
\item The applications of \texttt{BODY1} are not assumed to be done in any
  particular order: they can go from \texttt{v[0]} upwards to
  \texttt{v[length(v)-1]} or in the opposite direction.  Therefore,
  all applications of \texttt{BODY1} should be independent from each
  other. \mclcomm{That is at odds with the first item. }\stcomm{It maybe depends on the target language. Not sure about it.}
\end{itemize}

\begin{figure}
\centering
\parbox{0.8\linewidth}{
\begin{lstlisting}[style=cstyleRules]
join_assignments {
    pattern: {
        cstmts(s1);
        cexpr(v) = cexpr(e1);
        cstmts(s2);
        cexpr(v) = cexpr(e2);
        cstmts(s3);
    }
    condition: {
        no_write(cstmts(s2), {cexpr(v),cexpr(e1)});
        no_read(cstmts(s2),{cexpr(v)});
        pure(cexpr(e1));
        pure(cexpr(v));
    }
    generate: {
        cstmts(s1);
        cstmts(s2);
        cexpr(v) = subs(cexpr(e2),cexpr(v), cexpr(e1));
        cstmts(s3);
    }  
}
\end{lstlisting}}
 \caption{The \stml rule \textsc{JoinAssignments} in C-like syntax.}
 \label{fig:stml-rule-JoinAssignments}
\end{figure}

The \stml properties inferred from some high-level annotations are
shown in Table~\ref{tab:polca_stml_translation}.  Focusing on the
translation of \texttt{map}, the \stml annotations mean that:

\begin{itemize}
\item Iteration $i$-th reads from \texttt{v} in the position $i$-th
  (it actually reads from the set of positions \{$i+0$-th\}, since the
  set of offsets it reads from is \{0\}).
\item Iteration $i$-th writes on \texttt{w} in the position
  $i$-th. 
\item \texttt{v} and \texttt{w} have the same length.
\item \texttt{F} behaves as if it did not have side effects.
\item \texttt{F} is applied to \texttt{v} and \texttt{w} in the
  indexes ranging from $0$ to $length(\mathtt{v})$. 
\end{itemize}


Table~\ref{tab:polca_stml_translation} shows the \stml properties
inferred from
other high-level annotations (explained more in depth in~\cite{cluster2015}).
Fig.~\ref{lst:map_translation_example} shows the translation of the
code in Fig.~\ref{lst:polca_ann_example} into \stml.  All rules
used in the transformation in Fig.~\ref{fig:code_trans_seq} are shown
in Table~\ref{tab:transformations-math-1}, and the conditions they need
are described in Table~\ref{tab:basic-predicates}.
\mclcomm{I'm not sure to understand the symbols for the 5th and 6th
  ones?  Those which have $<$ and $-a[l]$?}

\begin{table*}[t]
   \begin{center}
     \begin{tabular}{l@{}}
           \begin{minipage}{6cm}
       \begin{lstlisting}
 for($l$=$e_{ini}$;$rel(l,e_{end})$;$mod(l)$) {$s_1$}
 for($l$=$e_{ini}$;$rel(l,e_{end})$;$mod(l)$) {$s_2$}
 \end{lstlisting}
      \end{minipage}
       \rw~~
       \begin{minipage}{7cm}
       \begin{lstlisting}
 for($l$=$e_{ini}$;$rel(l,e_{end})$;$mod(l)$) {$s_1$;$s_2$}
       \end{lstlisting}
       \end{minipage}\vspace{-2ex}\\
 	\when\ $rel~\pure$,
       \lstinline|$(s_1$;$s_2) \nowrites \{l, e_{ini}, e_{end}\}$|, $\writes(mod(l)) \subseteq \{l\},$
       $~ s_1 \nowritesexcarrays ~ s_2, s_2 \nowritesexcarrays ~ s_1, s_2 \nowritesarray ~ s_1$\\
       \hfill \textsc{(For-LoopFusion)}\\[1ex]
       \lstinline|$l$ += $e$;|
        \rw\ 
       \lstinline|$l$ = $l$ + $e$;| \\
       \when\ $l~\pure$
       \hfill \textsc{(AugAdditionAssign)}\\[1ex]
       \lstinline|$s_1$; $l$ = $e_1$; $s_2$; $l$ = $e_2$; $s_3$;| \rw\ 
       \lstinline|$s_1$; $s_2$; $l$ = $e_2[e_1/l]$; $s_3$;| \\ 
       \when\ $l, e_1~\pure, s_2 \nowrites\ \{l, e_1\}, s_2 \noreads\ l, s_2 \nowrites\ e_1$
      \hfill \textsc{(JoinAssignments)}\\[1ex]
      $f(g(e_1,e_3),g(e_2,e_3))$ \rw\ $g(f(e_1,e_2),e_3)$ \\ 
       \when\ $e_1, e_2, e_3~\pure, g~\distributesover\ f$
       \hfill \textsc{(UndoDistribute)}\\[1ex]
       \lstinline|for ($e_1$;$e_2$;$e_3$){$s_b$}| \rw\
       \lstinline|$l$ = $e_{inv}$; for ($e_1$;$e_2$;$e_3$){$s_b[l/e_{inv}])$}|\\
       \when\ $l~\fresh, e_{inv} ~\occurs~ s_b, e_{inv}~\pure, \{s_b, e_3, e_2 \}\nowrites e_{inv}$
       \hfill \textsc{(LoopInvCodeMotion)}\\[1ex]
    \end{tabular}
  \end{center}
   \caption{Source code transformations used in the
     example of Fig.~\ref{fig:code_trans_seq}.}
   \label{tab:transformations-math-1}
\end{table*}

\begin{table*}[t]
\begin{center}
\begin{tabular}{l@{~~~}p{0.75\linewidth}}
$s \nowrites\ l$ &
 statements $s$ do not write into location $l$: $l \notin \writes(s)$ \\
$s \noreads\ l$ &
 statements $s$ do not read the value in location $l$\\
$s_1 \nowrites\ s_2$ &
 statements $s_1$ do not write into any location read by $s_2$\\
 $s_1 \noreads\ s_2$ &
 statements $s_1$ do not read from any location written by $s_2$\\
 $s_1 \nowritesexcarrays\ s_2$ &
 same predicate as the previous one but not taking into account locations referred through arrays\\
 $s_1 \nowritesarray\ s_2$ &
 statements $s_1$ do not write into any previous location corresponding to an index array read by $s_2$\\
$e~\pure$ &
 expression $e$ is \emph{pure}, i.e.~does not have side effects nor
 writes any memory locations\\
 $\writes(s)$ &
 set of locations written by statements $s$.\\
{$g~\distributesover~f$}&
$\forall x, y, z .~ g(f(x,y),z) \approx f(g(x,z),g(y,z))$\\
$l~\fresh$ &
 $l$ is the location of a \emph{fresh} identifier, i.e.~does not clash with existing
 identifiers if introduced in a given program state\\
\end{tabular}
\end{center}
\caption{Predicates used to express conditions for the application
         transformation rules in Table \ref{tab:transformations-math-1}.}
\label{tab:basic-predicates}
\end{table*}

\begin{figure}

\begin{tabular}{c|c}
\noindent\begin{minipage}{.45\textwidth}
\begin{lstlisting}[basicstyle={\small\ttfamily}]
float c[N], v[N], a, b;

#pragma polca map BODY1 v c
#pragma stml reads v in {0}
#pragma stml writes c in {0}
#pragma stml same_length v c
#pragma stml pure BODY1
#pragma stml iteration_space 0 length(v)
#pragma stml iteration_independent
for(int i = 0; i < N; i++)
#pragma polca def BODY1
#pragma polca input v[i]
#pragma polca output c[i]
   c[i] = a*v[i];
\end{lstlisting}
\end{minipage} &

\noindent\begin{minipage}{.45\textwidth}
\begin{lstlisting}[basicstyle={\small\ttfamily}]
#pragma polca zipWith BODY2  v c c
#pragma stml reads v in {0} 
#pragma stml reads c in {0}
#pragma stml writes c in {0}
#pragma stml same_length v c
#pragma stml pure BODY2
#pragma stml iteration_space 0 length(v)
#pragma stml iteration_independent
for(int i = 0; i < N; i++)
#pragma polca def BODY2
#pragma polca input v[i]
#pragma polca input c[i]
#pragma polca output c[i]
   c[i] += b*v[i];
\end{lstlisting}
\end{minipage}
\end{tabular}
\caption{Translation of high-level annotations in Fig.~\ref{lst:polca_ann_example} into \stml.}
\label{lst:map_translation_example}
\end{figure}

\subsubsection{External Tools}

Besides the properties provided by the user,
external tools can automatically infer
additional properties, thereby
relieving users from writing many annotations to
capture low-level details.  These properties can be made
available to the transformation tool by writing them as \stml
annotations.
%
%
We are currently using Cetus~\cite{dave2009cetus} to automatically
produce \stml annotations.  Cetus is a compiler framework, written in
Java, to implement source-to-source transformations. We have modified
it to add some new analyses and to
output the properties it infers as \stml pragmas annotating the
input code.
%
%
If the annotations automatically inferred by external tools contradict
those provided by the user, the properties provided by the user are
preferred to those deduced from external tools, but a warning is
issued nonetheless.

\begin{table*}[t]
  \begin{center}
    \begin{tabular}{|  p{0.44\textwidth}  p{0.50\textwidth}  |}
      \hline
      \textbf{Function} & \textbf{Description}\\
      \hline
      \texttt{no\_write((S|[S]|E)$_\mathtt{1}$,} \texttt{(S|[S]|E)$_\mathtt{2}$)} 
      	&  True if \texttt{(S|[S]|E)$_\mathtt{1}$} does not write
      ~in any location read by \texttt{(S|[S]|E)$_\mathtt{2}$}.\\
      \hline
     \texttt{no\_write\_except\_arrays} 
    \texttt{\hspace*{2em}((S|[S]|E)$_\mathtt{1}$,(S|[S]|E)$_\mathtt{2}$,}\texttt{E)}        
      	&  As the previous condition, but not taking arrays accessed 
      ~using \texttt{E} into account.\\
      \hline
     \texttt{no\_write\_prev\_arrays} 
     \texttt{\hspace*{2em}((S|[S]|E)$_\mathtt{1}$}
      \texttt{(S|[S]|E)$_\mathtt{2}$,} \texttt{E)} 
          &  True if no array writes indexed using \texttt{E} in
            \texttt{(S|[S]|E)$_\mathtt{1}$} access previous locations
            to array reads indexed using  \texttt{E} in
            \texttt{(S|[S]|E)$_\mathtt{2}$}.
        \\ \hline
    \texttt{no\_read((S|[S]|E)$_\mathtt{1}$,} \texttt{(S|[S]|E)$_\mathtt{2}$)} 
      	&  True if \texttt{(S|[S]|E)$_\mathtt{1}$} does not read in
      any location written by \texttt{(S|[S]|E)$_\mathtt{2}$}.\\
      \hline
      \texttt{pure((S|[S]|E))} 
      	&  True if \texttt{(S|[S]|E)} does not write
	in any location.\\
      \hline
     \texttt{writes((S|[S]|E))} 
      	&  Locations written by \texttt{(S|[S]|E)}.\\ 
      \hline
      \texttt{distributes\_over(E$_\mathtt{1}$,E$_\mathtt{2}$)} 
      	&  True if operation \texttt{E$_\mathtt{1}$} distributes over operation \texttt{E$_\mathtt{2}$}.\\ 
      \hline
      \texttt{occurs\_in(E,(S|[S]|E))} 
      	&  True if expression \texttt{E} occurs in \texttt{(S|[S]|E)}.\\ 
      \hline
      \texttt{fresh\_var(E)} 
      	&  Indicates that \texttt{E} should be a new variable. \\ 
      \hline
      \texttt{is\_identity(E)} 
      	&  True if \texttt{E} is the identity.\\ 
     \hline
      \texttt{is\_assignment(E)} 
      	&  True if \texttt{E} is an assignment.\\ 
      \hline
      \texttt{is\_subseteq(E$_\mathtt{1}$,E$_\mathtt{2}$)} 
      	&  True if \texttt{E$_\mathtt{1}$} $\subseteq$ \texttt{E$_\mathtt{2}$}
	\\ 
      \hline
    \end{tabular} 
  \end{center}
  \caption{Rule language functions for the  \texttt{condition} section
    of a rule.}
  \label{tab:funscondition}
\end{table*}

\begin{table*}[t]
  \begin{center}
    \begin{tabular}{| p{0.37\textwidth}  p{0.57\textwidth} |}
      \hline
      \textbf{Function/Construction} & \textbf{Description}\\
      \hline
      \texttt{subs((S|[S]|E),E$_\mathtt{f}$,E$_\mathtt{t}$)} 
      	& Replace each occurrence of \texttt{E$_\mathtt{f}$} in
        \texttt{(S|[S]|E)} for \texttt{E$_\mathtt{t}$}.\\
      \hline
      \texttt{if\_then:\{E$_\mathtt{cond}$;} 
     \texttt{(S|[S]|E);\}}
       & If \texttt{E$_\mathtt{cond}$} is true,
        ~then generate \texttt{(S|[S]|E).}\\
     \hline
   \texttt{if\_then\_else:}\texttt{\{E$_\mathtt{cond}$;}&If \texttt{E$_\mathtt{cond}$} is true, 
then generate \texttt{(S|[S]|E)$_\mathtt{t}$}\\		
\texttt{~~(S|[S]|E)$_\mathtt{t}$;}\texttt{(S|[S]|E)$_\mathtt{e}$;\}} &
\hspace*{6.2em}else generate  \texttt{(S|[S]|E)$_\mathtt{e}$}.\\
      \hline
      \texttt{gen\_list:}  \texttt{\{[(S|[S]|E)];\}}
      	&  Each element in \texttt{[(S|[S]|E)]} 
       produces a different rule consequent.\\
      \hline
    \end{tabular} 
  \end{center}
  \caption{Rule language constructions and functions for \texttt{generate} rule section.}
  \label{tab:funsgenerate}
\end{table*}

\subsection{Rules in STML}
\label{sec:stml_rules}

Let us see one example:  Fig.~\ref{fig:stml-rule-JoinAssignments}
shows the \stml version of rule
\textsc{JoinAssignments}.
Rules can be applied when the
code being transformed matches 
the \texttt{pattern} section and fulfills the \texttt{con\-di\-tion}
section.  When the rule is activated, code is generated according to
the template in the \texttt{generate} section, where expressions
matched in the \texttt{pattern} are replaced in the \texttt{generate}d
code.  In this case one assignment is removed by propagating the
expression in its \emph{right hand side} (RHS).


\stml uses tagged meta-variables to match components of the initial
code and specify which kind 
of component is matched.  For example, a meta-variable
\texttt{v} can be 
tagged as \texttt{cexpr(v)} to denote that it can only match an
expression, \texttt{cstmt(v)} for a statement, or \texttt{cstmts(v)}
for a sequence of statements.
%
%
In Fig.~\ref{fig:stml-rule-JoinAssignments}, \texttt{s1}, \texttt{s2}
and \texttt{s3} should be (sequences of) statements, and \texttt{e1},
\texttt{e2} and \texttt{v} are expressions.

Additional conditions and primitives (Tables~\ref{tab:funscondition}
and~\ref{tab:funsgenerate}) help write descriptive rules that can at
the same time be sound.
In these tables, \texttt{E} represents an expression, \texttt{S}
represents a statement and \texttt{[S]} represents a sequence of
statements.  The function
\texttt{bin\_oper(E$_\mathtt{op}$,E$_\mathtt{l}$,E$_\mathtt{r}$)}
matches or generates a binary operation \texttt{(E$_\mathtt{l}$
  E$_\mathtt{op}$ E$_\mathtt{r}$)} and can be used in the sections
\texttt{pattern} and \texttt{generate}.  The section \texttt{generate}
can also state, using \texttt{\#pragma}s, new properties that hold in
the resulting code.

\subsection{Rule Selection}
\label{sec:trans_ml_inter}

\begin{enclosedreminders}
  \begin{quote}
  \textbf{Where do we put this?  It's a partial repetition}

  Deciding whether a given rule is applicable or not depends on
  whether rule conditions are met or not.  In practice, it is often
  the case that the tool does not have information enough to decide
  whether a property holds or not, and therefore it cannot be decided
  whether
%
%
a rule that requires that property can be applied or not. Thus,
the system distinguishes between definitely ``applicable'',
``definitely not applicable'' and ``probably (not) applicable''
transformation steps.  For all applicable rules, deciding which one
should be chosen
should be based on whether that rule can 
contribute to an eventual improvement of the performance of the final code with respect to the original one. 
\end{quote}
\end{enclosedreminders}

In most cases, several (often many) rules can be safely applied at
multiple code points in every step of the rewriting process.  Deciding
which rule has to be fired should be ultimately decided based on
whether that rule contributes to an eventual increase in performance.
%
%
As mentioned before, we currently provide two ways to perform rule selection: a
human-driven one and an interface to  communicate with external
tools.

%


\subsubsection{Interactive Rule Selection}
\label{sec:interactive-selection}

An interface to make interactive transformations possible is
available: the user is presented with the rules that can be applied at
some point together with the piece of code before and after applying
some selected rule (using auxiliary programs, such as~\cite{meld}, to
clearly show the differences).  This is
useful to refine/debug rules or
to perform general-purpose refactoring,
which may or not be related to improving performance or adapting code to a
given platform.

\subsubsection{Oracle-Based Rule Selection}
\label{sec:oracle-selection}

In our experience, manual rule selection is very fine-grained and in
general not scalable, and using it is not realistic even for
medium-sized programs.
Therefore, mechanizing as much
as possible this process is a must, keeping in mind that our goal is that the
final code has to improve the original code.
A straightforward possibility is to select at each step the
rule 
that improves more some metric.  However, this may make the search to be
trapped in local minima.  In our experience, it is often necessary to
apply transformations that temporarily reduce the quality of the code
because they enable the application of further transformations.

A possibility to work around this problem is to explore a bounded
neighborhood.  The size of the bounded
region needs to be decided, since taking too few steps would not make
it possible to leave a local minimum.  Given that in our experience the
number of rules that can be applied in most states is high (typically
in the order of the tens), increasing the diameter of the boundary to
be explored can cause an exponential explosion in the number of states
to be evaluated. This would happen even considering some improvements
such as partial order reduction for pairs of commutative rules.

Therefore, we need a mechanism that can make local decisions taking
into account global strategies --- i.e., a procedure able to select a
rule under the knowledge that it is part of a sequence of rule
applications
that improves code performance for a given platform.  We are exploring
the use of machine learning techniques based on
%
\emph{reinforcement learning}~\cite{vigueras16:learning-prole}.
From the point of view of the transformation engine, the selection
tool works as an \emph{oracle} that, given a code configuration and a
set of applicable rules, returns which rule should be applied.  We
will describe now an abstract interface to an external rule
selector, which can be applied not only to the current oracle, but
to other similar external oracles.   



\newcommand{\AppRules}{\mathit{AppRules}}
\newcommand{\Code}{\mathit{Code}}
\newcommand{\Rule}{\mathit{Rule}}
\newcommand{\Pos}{\mathit{Pos}}
\newcommand{\Trans}{\mathit{Trans}}

\newcommand{\SelectRule}{\mathit{SelectRule}}
\newcommand{\IsFinal}{\mathit{IsFinal}}
\newcommand{\Boolean}{\mathit{Boolean}}

\newcommand{\NewCode}{\mathit{NewCode}}
\newcommand{\rules}{\mathit{rls}}
\newcommand{\AllRules}{\mathit{AllRules}}

\begin{figure}
  \begin{minipage}[t]{0.48\linewidth}
$\AppRules(\Code) \rightarrow \{(\Rule, \Pos)\}$ \\
$\Trans(\Code_i, \Rule, \Pos) \rightarrow \Code_o$
    \caption{Functions provided by the transformation tool.}
    \label{fig:trans-tool-functions}
  \end{minipage}
\hfill
  \begin{minipage}[t]{0.5\linewidth}
$\SelectRule(\{(\Code_i, \{\Rule_i\})\}) \rightarrow (\Code_o, \Rule_o)$ \\
$\IsFinal(\Code) \rightarrow \Boolean$
    \caption{Functions provided by the oracle.}
    \label{fig:oracle-functions}
  \end{minipage}
\end{figure}


The interface of the transformation tool
(Fig.~\ref{fig:trans-tool-functions}) is composed by functions
$\AppRules$ and $\Trans$.  Function $\AppRules$ determines the
possible transformations applicable to a given code and returns, for a
given input $\Code$, a set of tuples containing each a rule name
$\Rule$ and the code position $\Pos$ where it can be applied (e.g.,
the identifier of a node in the AST).  Function $\Trans$ applies rule
$\Rule$ to code $\Code_i$ at position $\Pos$ and returns the resulting
code $\Code_o$ after applying the transformation.


The API from the external tool (Fig.~\ref{fig:oracle-functions})
includes operations to
decide which rule has to be applied and whether the search should
stop.  Function $\SelectRule$ receives a set of safe possibilities,
each of them composed of a code fragment and a set of rules that can be
applied to it, and returns one of the input code fragments and the
rule that should be applied to it.
Function $\IsFinal$ is used to know whether a given code $\Code$ is
considered ready for translation or not.


The function that defines the interaction between the transformation
engine and the external oracle is
$\NewCode$ (Fig.~\ref{fig:interaction}),
%
which receives an initial
$\Code_i$ and a set of rules and returns (a) 
$\Code_o$ which results from applying one of the rules from
$\{\Rule_i\}$ to $\Code_i$, and (b) 
$\Rule_o$ that should be applied in the next transformation step,
i.e., the next time $\NewCode$ is invoked with
$\Code_o$.  The rationale is that the first time
$\NewCode$ is called, it receives all the applicable rules as
candidates to be applied, but after this first application
$\{\Rule_i\}$ is always a singleton. 
$\NewCode$ is called repeatedly until the transformation generates
a code for which $\IsFinal$ returns true. 

\begin{figure}
  \begin{minipage}[t]{0.64\linewidth}
\textbf{Header} \\
$\NewCode(\Code_i, \{\Rule_i\}) \rightarrow (\Code_o, \Rule_o)$  \\[1ex]
\textbf{Definition}\\
 $\NewCode(c, \rules) = $
$\SelectRule(\{(c', \{r'~|~(r',\_) \in \AppRules(c')\})$\\
$~~~ ~~|~ c' \in \{\Trans(c, r, p)~|~ (r, p) \in \AppRules(c), r \in \rules \} ~\})$
  \end{minipage}
\hfill
  \begin{minipage}[t]{0.35\linewidth}
\textbf{Complete derivation}\\
$\NewCode(c_0, \AllRules) \rightarrow^* (c_n,r_n)$\\
\hspace*{0.3em}
    when $\IsFinal(c_n)$ and $\forall i, 0 < i < n.\\
    (c_i, r_i) = \NewCode(c_{i -1}, \{r_{i -1}\})$\\
\hspace*{0.3em}
    when $\neg \IsFinal(c_i)$
  \end{minipage}
  \caption{Interaction between the transformation and the oracle interface.}
  \label{fig:interaction}
\end{figure}


This approach makes it unnecessary for the external oracle to consider
code positions where a transformation can be applied, since that choice is
implicit in the selection of a candidate code between all possible
code versions obtained using a single input rule.  Furthermore, by
selecting the next rule to be applied, it takes the control of the
next step of the transformation. The key here is the function
$\SelectRule$:  given  inputs $\Code_i$ and $\Rule_i$,
$\SelectRule$ selects
a resulting code between all the codes that can be generated from
$Code_i$ using $Rule_i$. The size of the set received by function
$SelectRule$ corresponds to the total number of positions where
$Rule_i$ can be applied. In this way, $SelectRule$ is implicitly
selecting a position.

\section{Producing Code for Heterogeneous Systems}
\label{sec:translation}

In the second phase of the tool (Fig.~\ref{fig:ana-trans-tool}), code
for a given platform is produced starting from the result of the
transformation process.  The destination platform of a fragment of
code can be specified using annotations that make this explicit.
%
This information helps the tool decide what transformations should be
applied and when the code is ready for translation.  


The translation to code for a given architecture is in most cases
straightforward as it needs only to introduce the ``idioms'' necessary
for the architecture or to perform a syntactical translation.  As a
consequence, there is no search or decision process: for each input
code given to the translation, there is only one output code
that is obtained via predefined transformations or glue code
injection.

%

Some translations need specific information: for instance,
knowing if a statement is performing I/O is necessary when translating
to MPI, because  executing this operation might need to be done in a
single thread.  It is often the case that this can be deduced by
syntactical inspection, but in other cases (e.g., if the operation is
part of a library function) it may need explicit annotations.



\section{Implementation Notes}
\label{sec:implementation}


The transformation phase, which obtains C code that could be easily
translated into the source language for the destination platform, is a
key part of the tool.  As a large part of the system was experimental
(including the definition of the language, the properties, the
generation of the final code, and the search / rule selection
procedures), we needed a flexible and expressive implementation
platform.   We decided to use a declarative language and
implement the tool in Haskell.  Parsing the input code is done by
means of the \texttt{Language.C}~\cite{LanguageC} library which
returns the AST as a data structure that is easy to manipulate.  In
particular, we used the Haskell facilities to deal with generic data
structures through the \emph{Scrap Your Boilerplate} (SYB)
library~\cite{DataGenerics}.  This allows us to easily extract
information from the AST or modify it with a generic traversal of the
whole structure.

The rules themselves are written in a subset of C and are
parsed using \texttt{Language.C}.  After reading these rules in, they
are automatically compiled into Haskell code (contained in the file
\texttt{Rules.hs} ---see Fig.~\ref{fig:ana-trans-tool}) that
performs the traversal and (when applicable) the transformation of the
AST.  This module is loaded with the rest of the tool,
therefore avoiding the extra overhead of interpreting the rules.  

When it comes to rule compilation, \stml rules can be divided into two
classes: those that operate at the expression level
and those that can manipulate both expressions and sequences of
statements.  In the latter case, sequences of statements
(\texttt{cstmts}) of an unknown size have to be considered: for
example, in Fig.~\ref{fig:stml-rule-JoinAssignments}, \texttt{s1},
\texttt{s2}, and \texttt{s3} can be sequences of any number of
statements (including the empty sequence), and the rule has to try all
the possibilities to determine if there is a match that meets the
rule conditions. For this, Haskell code that explicitly performs an
AST traversal needs to be generated.  Expressions, on the other hand,
are syntactically bound and the translation of the rule is much
easier.

When generating Haskell code, the rule sections (\texttt{pattern},
\texttt{condition}, \texttt{generate}, \texttt{assert}) generate the 
corresponding LHS's, guards, and RHS's of a Haskell function.
If the conditions to apply a rule are met, the result is returned in a
triplet \ihaskell{(rule\_name, old\_code, new\_code)} where the two
last components are, respectively, the matched and transformed
sections of the AST. 
\mclcomm{has this changed?}  
\stcomm{Not. It is still the same. Properties are attached in the AST, therefore there is not need to modify this.}  
Note that \ihaskell{new\_code} may contain new properties if the
\texttt{generate} section of the rule defines them.


The tool is divided into four main modules: 

\begin{itemize}

\item \texttt{\textbf{Main.hs}} 
  implements the main  workflow of the tool: it calls the parser on the input C
  code to build the AST, links the pragmas to the AST, executes the 
  transformation sequence (interactively or automatically)
  and outputs the transformed code.

\item \texttt{\textbf{PragmaLib.hs}} reads pragmas and links them to
  their corresponding node in the AST.  It also restores or injects
  pragmas in the transformed code.

\item \texttt{\textbf{Rul2Has.hs}} translates \stml rules (stored in
  an external file) into Haskell functions that actually perform the
  AST manipulation.  It also reads and loads \stml rules as an AST and
  generates the corresponding Haskell code in the
  \texttt{\textbf{Rules.hs}} file.

 
\item \texttt{\textbf{RulesLib.hs}} contains supporting code used by
  \texttt{Rules.hs} to identify whether some \stml rule is or not
  applicable (e.g., there is matching code, the preconditions hold,
  etc.) and to execute the implementation of the rule (including AST
  traversal, transformation, \ldots).
\end{itemize}



\section{Conclusion}
\label{sec:conclusions}

We have presented a transformation toolchain that uses semantic
information, in the form of user- or machine-provided annotations, to
produce code for different platforms.  It has two clearly separated
phases: a source-to-source transformation that generates code with
the style appropriate for the destination architecture and a
translation from that code to the one used in the specific platform.

We have focused until now in the initial phase, which included the
specification of a DSL (\stml) to define rules and code properties, a
translator from this language into Haskell, a complete engine to work with
these rules, and an interface to interact with external oracles (such
as a reinforcement learning tool that we are developing) to guide the
transformation.

The translation phase is still in a preliminary stage. However, and
while it is able to translate some input code, it needs to be improved
in order to support a wider range of programs.  
We have compared, using several metrics, the code obtained using our
tool and the corresponding initial code and the results are
encouraging. 

As said before, we have started the development of external
(automated) oracles to guide the transformation process. Initial
results using an oracle based on reinforcement (machine)
learning~\cite{vigueras16:learning-prole} are very encouraging.  The
possibility of using other techniques such as partial order reduction
to prune the search space is still open to investigation.

We plan to improve the usability of the \stml language
and continue modifying Cetus to automatically obtain more
advanced\,/\,specific properties, and we are integrating profiling
techniques in the process to make  evaluating the whole
transformation system and giving feedback on it easier.
Simultaneously, we are investigating other analysis tools that can
be used to derive more precise properties.  Many of these
properties are related to data dependencies and pointer behavior.  We
are considering, on one hand, tools like
PLuTo~\cite{Bondhugula:2008} and PET~\cite{poly_pet} (two polytope
model-based analysis tools) or the dependency analyzers for the
Clang\,/\,LLVM compiler.  However, since they fall short to derive
dependencies (e.g., alias analysis) in code with pointers, we are also
considering tools based on separation
logic~\cite{DBLP:conf/lics/Reynolds02} such
as VeriFast~\cite{DBLP:conf/nfm/JacobsSPVPP11}
that can reason on dynamically-allocated mutable structures.

\bibliographystyle{eptcs}
\bibliography{../../BiBTeX/hpc_transformations,%
              ../../BiBTeX/polca_refs,%
              ../../BiBTeX/polca_deliverables,%
              ../../BiBTeX/c_a_t}


%
%


\end{document}